\documentclass[aps,prl,twocolumn,showpacs,superscriptaddress]{revtex4}
\usepackage{graphicx}
\usepackage{dcolumn}
\usepackage{bm}

\def\ii{{\rm i}}

\begin{document}
\title{Tunneling mechanism of light transmission through metallic films}
\author{F.~J.~Garc\'{\i}a~de~Abajo}
\affiliation{Unidad de F\'{\i}sica de Materiales CSIC-UPV/EHU
Aptdo. 1072, 20080 San Sebastian, Spain} \affiliation{Donostia
International Physics Center (DIPC), Aptdo. 1072, 20080 San
Sebastian, Spain}
\author{G.~G\'{o}mez-Santos}
\affiliation{Departamento de F\'{\i}sica de la Materia Condensada,
Facultad de Ciencias, Universidad Aut\'{o}noma de Madrid,
Cantoblanco, 28049 Madrid, Spain}
\author{L.~A.~Blanco}
\affiliation{Donostia International Physics Center (DIPC), Aptdo.
1072, 20080 San Sebastian, Spain}
\author{A.~G.~Borisov}
\affiliation{Laboratoire des Collisions Atomiques et
Mol\'{e}culaires, Universit\'{e} Paris-Sud, B\^{a}timent 351,
F-91405 Orsay Cedex, France}
\author{S.~V.~Shabanov}
\affiliation{Department of Mathematics, University of Florida, FL
32611, USA}

\date{\today}

\begin{abstract}
A mechanism of light transmission through metallic films is
proposed, assisted by tunnelling between resonating buried
dielectric inclusions. This is illustrated by arrays of Si spheres
embedded in Ag. Strong transmission peaks are observed near the
Mie resonances of the spheres. The interaction among various
planes of spheres and interference effects between these
resonances and the surface plasmons of Ag lead to mixing and
splitting of the resonances. Transmission is proved to be limited
only by absorption. For small spheres, the effective dielectric
constant can be tuned to values close to unity and a method is
proposed to turn the resulting materials invisible.
\end{abstract}

\pacs{78.66.Bz,42.25.Fx,73.20.Mf,42.79.Ci}

\maketitle

Substantial efforts have been placed on studying light
transmission through perforated metallic films since the discovery
by Ebbesen {\it et al.} \cite{ELG98} of extraordinary transmission
through sub-wavelength periodic hole arrays, leading to such
remarkable effects as giant funnelling of microwave radiation
through metallic micrometer slits \cite{T01}.
Two distinct mechanisms have been identified that contribute to
enhance light transmission: (i) dynamical scattering assisted by
surface plasmons (SPs) and (ii) resonant transmission by coupling
to propagating modes. (i) Isolated sub-wavelength holes can only
support evanescent modes, but dynamical diffraction assisted by
SPs in periodic hole arrays lead to transmission maxima that can
beat the severe $1/\lambda^4$ dependence on wavelength $\lambda$
of the transmission derived by Bethe for single apertures
\cite{B1944}, although there is still some controversy regarding
the relative relevance of SPs and dynamical diffraction
\cite{T99,PNE00,MGL01,SVV03}.
(ii) Slits differ from holes in that the former, however small,
support at least one transmission mode that can couple resonantly
to external light to transmit several orders of magnitude more
light than impinging directly on the slit aperture \cite{T01}.

In this Letter we investigate yet another mechanism of enhanced
transmission based upon hopping between dielectric inclusions that
can sustain localized resonances inside metallic films. We
illustrate this concept by studying light transmission through
metallic Ag films that contain spherical Si inclusions arranged in
square-lattice layers, which belongs to the family of periodic
arrays of nanoresonators coupled via spatially homogeneous
waveguides \cite{Kuhl}. The transmission is shown to take large
values near the Mie resonances of the spheres, and it is only
limited by metal absorption.

\begin{figure}
\centerline{\scalebox{0.35}{\includegraphics{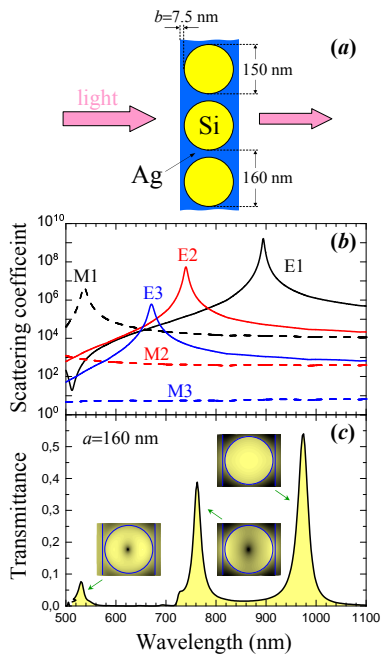}}}
\caption{\label{Fig1} {\bf (a)} Transversal section of a Ag film
containing a buried square planar array of 150-nm Si spheres in
square lattice configuration. {\bf (b)} Square modulus of the Mie
scattering coefficients of single 150-nm Si spheres inside bulk Ag
for different multipoles ($l=1,2,3$) and polarization modes
[electric (E) and magnetic (M)] as a function of light wavelength in
vacuum. {\bf (c)} Absolute transmittance through the film of (a)
under normal incidence for a lattice constant $a=160$ nm. The insets
show maps of the electric field strength in the slab in logarithmic
scale for circularly polarized incident light at the peak
wavelengths.}
\end{figure}

We first consider a single layer of inclusions [Fig.\
\ref{Fig1}(a)]. Maxwell's equations are solved rigourously for
this geometry using a layer-by-layer version of the KKR
multiple-scattering method \cite{SYM98}, with lattice sums
performed in real space within the metal. The complex,
frequency-dependent dielectric functions of Ag and Si are taken
from optical data \cite{P1985}.

The resonant modes of the Si spheres can be visualized through
their Mie scattering coefficients inside bulk metal \cite{J1975},
showing prominent features in the visible and near-IR domains for
a sphere radius $R=75$ nm [Fig.\ \ref{Fig1}(b)]. The corresponding
transmittance spectrum exhibits peaks near those resonances [Fig.\
\ref{Fig1}(c), at normal incidence] that can exceed a transmission
$T=50\%$ at wavelengths near $\lambda=975$ nm, well above the
$T=13\%$ value expected for a homogeneous Ag film of thickness
$2b=15$ nm (i.e., the combined thickness of the thin burying
layers at either sides of the spheres). Smaller separations
between the sphere and the film boundaries lead to even higher
transmission (e.g., $T=70\%$ for $b=2.5$ nm). The transmission
peaks are red-shifted with respect to the resonances in bulk Ag as
a result of interaction between sphere modes and vacuum modes.
However, the relevant multipoles of the inclusions and the
polarization symmetry associated to those peaks can be well
ascribed by comparing Figs.\ \ref{Fig1}(b) and (c).

\begin{figure}
\centerline{\scalebox{0.35}{\includegraphics{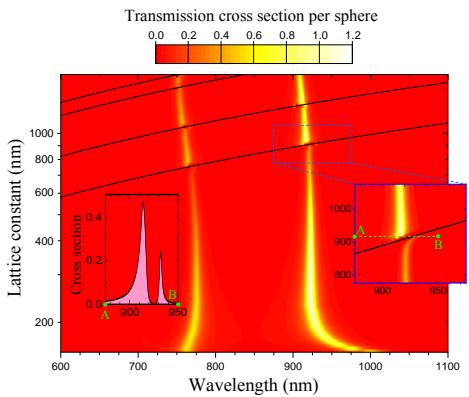}}}
\caption{\label{Fig2} Transmittance through the film described in
Fig.\ \ref{Fig1}(a) under normal incidence as a function of
wavelength and lattice constant. The transmittance is normalized to
the projected area of the spheres. The solid curves are given by the
condition that the momentum of the planar surface plasmon of Ag
matches some reciprocal lattice vector of the sphere lattice. The
right inset is a blow up of an interference region and the left
inset shows the cross section along the segment AB.}
\end{figure}

The interaction of spherical inclusions along the plane can be
important at small separations between the spheres. This is
illustrated in Fig.\ \ref{Fig2}, which shows a contour plot of the
transmittance $T$ as a function of lattice constant $a$ and
wavelength. $T$ has been multiplied by $a^2/\pi R^2$ to obtain
transmission cross sections per sphere
The $E1$ and $E2$ Mie resonances of Fig.\ \ref{Fig1}(b) show up as
two lines of transmission maxima relatively independent of $a$.
However, these two modes interact and repel each other when the
spheres are near touching each other (bottom part of the figure).

The SPs of the planar Ag interfaces cannot couple directly to
external propagating modes. However, the presence of Si inclusions
provides a source of momentum that makes this coupling possible.
The coupling increases when the interaction between lattice sites
(spheres) via SPs is constructive, which at normal incidence (all
lattice sites are in phase) leads to the kinematical condition
that SP momenta that match reciprocal lattice vectors of the
square lattice of spheres are favored, that is,
$\lambda\sqrt{n^2+m^2}=a\sqrt{\epsilon_{\rm Ag}/(\epsilon_{\rm
Ag}+1)}$ \cite{ELG98}, where the surface plasmon dispersion
relation has been invoked. The solid curves in Fig.\ \ref{Fig2}
are obtained from this equation for different combinations of
integers $n$ and $m$ and they would lie near regions of
transmission enhancement in hole arrays \cite{ELG98}, but that
transmission mechanism is marginal for our buried structures.
Instead, a rich transmission structure shows up when the noted
kinematical condition (solid curves) is at the same wavelength as
a Mie resonance [see right inset of Fig.\ \ref{Fig2}]: the
discrete, localized Mie modes couple to the continuum of SPs to
yield a transmission spectrum exhibiting a point of zero
transmission, as it is well known from Fano's theory \cite{F1961}
(see left inset of Fig.\ \ref{Fig2}).

\begin{figure}
\centerline{\scalebox{0.35}{\includegraphics{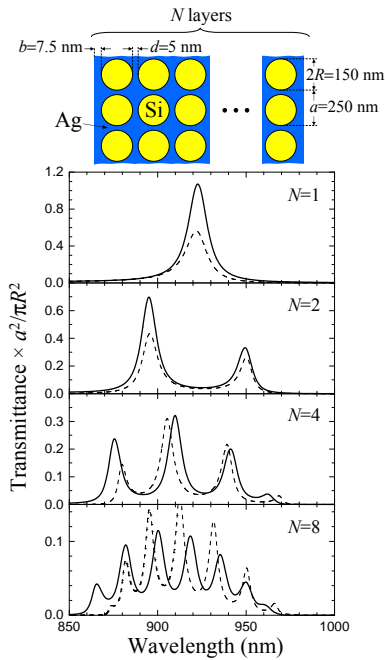}}}
\caption{\label{Fig3} Transmittance through Ag films containing
different numbers $N$ of buried planar arrays of Si spheres of
radius $R=75$ nm under normal incidence conditions. The planar unit
cell is a square of side $a=250$ nm, which prevents interaction
between spheres within each plane. Solid curves are exact solutions
of Maxwell's equations. Broken curves are obtained from the
tight-binding model explained in the text [Eq.\ (\ref{TT})].}
\end{figure}

Several layers of spheres can interact in a single film producing
splitting and shifting of the Mie resonances, as shown in Fig.\
\ref{Fig3}. The main 975-nm feature of Fig.\ \ref{Fig1}(c) is
subdivided here into a series of $N$ peaks in a film containing
$N$ layers. The coupling between Mie modes at consecutive layers
is the same as in previously introduced photonic molecules that
arise from the interaction of Mie modes in neighboring dielectric
particles \cite{MTM99}. The tight-binding approach used in that
context becomes a natural tool to understand the hopping of light
between contiguous spheres in our case. In this spirit, an analogy
with electron transmission between localized states in quantum
systems can be established by considering a simple tight-binding
model \cite{E83} in which a resonant mode trapped in each layer is
associated to a state $|i\rangle$ (with $i=1,2,...,N$). The
dynamics is governed by the Hamiltonian
\begin{eqnarray}
  H&=&(\omega_b+\ii\Sigma_s) (|1\rangle\langle 1| + |N\rangle\langle N|)
    + \sum_{i=2}^{i=N-1} (\omega_b+\ii\Sigma_b ) |i\rangle\langle i|
  \nonumber \\
     &+&  \sum_{i=1}^{i=N-1} (V |i\rangle\langle i+1| + h.c.),
  \nonumber
\end{eqnarray}
where $\omega_b$ and $\Sigma_b$ are the energy and width of the
resonances, and $V$ is the hopping amplitude, which expresses the
magnitude of the interaction between spheres. For internal layers
the width $\Sigma_b$ accounts for dissipation into bulk metal. The
states $|1\rangle$ and $|N\rangle$ interact with the continuum of
light modes on the left and right hand sides of the film,
respectively, described by frequency dependent self-energy
$\Sigma_s(\omega)$. The transmittance in the model is defined by
the squared absolute value of the transition amplitude and can be
written as \cite{E83}
\begin{eqnarray}
  T=4 |\Sigma_s(\omega) g_{1,N}(\omega+\ii 0^+)|^2,
  \label{TT}
\end{eqnarray}
where $g_{1,N}(z)=\langle 1|(z-H)^{-1}|N\rangle$. The parameters
of the Hamiltonian have been chosen to produce the best fit to the
transmittance obtained from numerical solution of Maxwell's
equations: $\Sigma_b=-0.002 \omega_b$, $V=0.03 \omega_b$, and a
linear function coupling to vacuum given by
$\Sigma_s(\omega)=-0.005\omega_b - 0.03 (\omega-\omega_b)$. Note
that radiative coupling via non-vanishing ${\rm Im}\{V\}$ can
produce changes in the line shapes \cite{fff} that lie beyond the
scope of this work.

The results derived from Eq.\ (\ref{TT}) are shown in Fig.\
\ref{Fig3} as dashed curves. Characteristic features of the exact
transmittance (solid curves) are well reproduced, including peak
splitting, peak asymmetry coming from the energy dependence of
$\Sigma_s$, and narrowing of the resonances with increasing number
of layers. This model corroborates that the main mechanism
producing transmission is tunnelling between electromagnetic modes
trapped in each layer.

\begin{figure}
\centerline{\scalebox{0.35}{\includegraphics{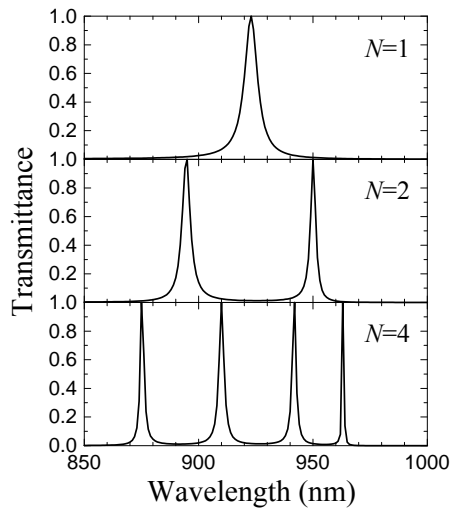}}}
\caption{\label{Fig4} Absolute transmittance under the same
conditions as in Fig.\ \ref{Fig3}, but setting the imaginary part of
the dielectric constant of Ag to zero.
}
\end{figure}

The dielectric function of Ag near the main resonance of Fig.\
\ref{Fig3} is essentially negative ($\epsilon_{\rm
Ag}=-49+0.6\ii$), so that coupling among the spheres and between
the spheres and the film planar surfaces is mediated by evanescent
modes. However, the small imaginary part of $\epsilon_{\rm Ag}$
plays a leading role in reducing the absolute transmittance (${\rm
Im}\{\epsilon_{\rm Si}\}\ll 1$ \cite{P1985}). This is made clear
in Fig.\ \ref{Fig4}, which has been calculated under the same
conditions as in Fig.\ \ref{Fig3} but setting the imaginary part
of $\epsilon_{\rm Ag}$ to a positive infinitesimal. All resonances
in the figure reach a maximum of 100\% transmission, a fact that
can be rigorously explained following simple arguments based upon
the change by $\pi$ of the phase of reflected and transmitted
resonant components of the field across the resonance
\cite{hole2}.



A Mie cavity mode inside a lossless bulk metal must have infinite
lifetime because light can neither scape through nor be absorbed
by the metal. Therefore, the width of the resonance in the
lossless film for $N=1$ (6.72 nm FWHM, as obtained from Fig.\
\ref{Fig4}) has to originate in the coupling of the $E1$ mode with
the continuum of free states on either side of the film (leakage
into the vacuum). For multilayer systems (e.g., $N=2$), as a
consequence of trace preservation in interacting systems as
compared to the non-interacting ones, this broadening has to
remain the same (indeed, the sum of the FWHM in the two peaks for
$N=2$ is 6.72 nm as well). This {\it leakage} contribution to peak
broadening adds up to the intrinsic width of the Mie resonance in
the lossy bulk metal [4.53 nm FWHM as obtained from the E1 peak in
Fig.\ \ref{Fig1}(b)], yielding a total width of 11.25 nm, which
compares well with the value of 11.83 nm observed for the actual
lossy metallic film ($N=1$ solid curve in Fig.\ \ref{Fig3}).

\begin{figure}
\centerline{\scalebox{0.35}{\includegraphics{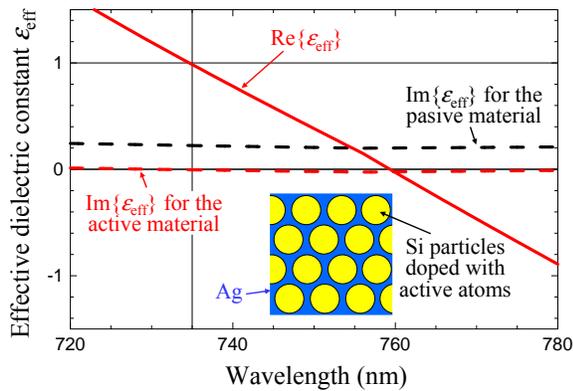}}}
\caption{\label{Fig5} Effective dielectric function of an fcc array
of Si spherical inclusions in Ag with a filling fraction of Si of
50\% (solid curves). The effective dielectric function is modified
when Si is doped with active centers (broken curves), described by
adding a negative imaginary part ($-0.252 \ii$) to the its
dielectric function. The resulting material is invisible
($\epsilon_{\rm eff}=1$) at a wavelength of 735 nm.}
\end{figure}

Therefore, our metallic film would be invisible to normally
incident radiation at specific resonant wavelengths if the metal
was free of absorption. The question arises whether the same
mechanism of hopping between dielectric inclusions in metals can
be utilized to fabricate solid structures that are invisible to
light incident with any angle at some wavelength, or in other
words, whether a medium exists exhibiting effective dielectric
function $\epsilon_{\rm eff}=1$ (and magnetic response $\mu_{\rm
eff}=1$ as well, of course). To explore this possibility, we have
represented $\epsilon_{\rm eff}$ in Fig.\ \ref{Fig5} for a
composite material containing 50\% of Si inside spherical
inclusions, which are arranged within Ag in an fcc lattice and
separated by distances much shorter than the wavelength.
$\epsilon_{\rm eff}$ is obtained using Maxwell-Garnett theory
\cite{M1904}, and we have found qualitatively similar results from
rigorous solution of Maxwell's equations for the reflectivity of a
surface of the composite material as compared to Fresnel equations
for an equivalent homogeneous medium. There is a wavelength at
which the real part of $\epsilon_{\rm eff}$ is 1, but its
imaginary part takes a small but non-negligible value.

We propose to dope Si in such a structure with optically-active
atoms or molecules that are capable of sustaining an inversion of
population under excitation by light of a different wavelength or
by electric discharge. The resulting lasing activity can be well
represented by a negative imaginary part contributing to the
dielectric function of doped Si \cite{KGZ01}, whose value can be
controlled by the doping dose or, in 4-level active atoms, also
via the pumping intensity. In particular, when this imaginary
contribution to $\epsilon_{\rm Si}$ is taken as $-0.252{\rm i}$
then $\epsilon_{\rm eff}=1$ at 735 nm (the imaginary part of
$\epsilon_{\rm eff}$ cancels out exactly), and the material
becomes invisible at that wavelength.


In conclusion, we have shown that light transmission through
metallic films can be enhanced by coupling to resonances localized
in dielectric inclusions, paving the way towards driving light
deep inside metals. Light circuits based upon this mechanism can
be envisioned. When the inclusions are small compared to the
wavelength, the resulting metamaterial can be made invisible
(i.e., $\epsilon_{\rm eff}=1$) if metal absorption is compensated
by doping the dielectric with active atoms under inversion
population conditions. We have proposed a specific design for such
invisible material at wavelengths in the visible, but structures
insensitive to microwaves should be also realizable due to smaller
absorption in this range and using reradiating active elements in
the millimeter scale.





\end{document}